\documentclass[twocolumn]{aastex631}
\usepackage{textcomp}
\usepackage{amssymb}
\usepackage{rotating}

\newcommand{\fermi}{\textit{Fermi}}
\newcommand{\gr}{$\gamma$-ray}
\newcommand{\grs}{$\gamma$-rays}

\begin{document}

\title{Identifying the Gamma-ray Emission of the Nearby Galaxy M83}

\author{Yi Xing}

\affil{Key Laboratory for Research in Galaxies and Cosmology, Shanghai Astronomical Observatory, Chinese Academy of Sciences, 80 Nandan Road, Shanghai 200030, China; yixing@shao.ac.cn; wangzx20@ynu.edu.cn} 

\author{Zhongxiang Wang}
\affil{Department of Astronomy, School of Physics and Astronomy, 
Yunnan University, Kunming 650091, China}
\affil{Key Laboratory for Research in Galaxies and Cosmology, Shanghai Astronomical Observatory, Chinese Academy of Sciences, 80 Nandan Road, Shanghai 200030, China; yixing@shao.ac.cn; wangzx20@ynu.edu.cn}

\begin{abstract}
We report on the detection of a \gr\ source at the position of the nearby 
	star-forming galaxy (SFG) M83, which is found from our analysis of
	14 years of the data obtained with the Large Area Telescope (LAT) 
	on-board {\it Fermi Gamma-ray Space Telescope (Fermi)}. 
	The source is weakly detected, with a significance of $\sim 5\sigma$,
	and its emission can be described with an exponentially cutoff power 
	law. At a distance of 4.61\,Mpc, the source's \gr\ luminosity is
	$\sim 1.4\times 10^{39}$\,erg\,s$^{-1}$, roughly along the correlation
	line between the \gr\ and IR luminosities determined for nearby
	SFGs. Because of the weak detection, the source 
	spectrum can not be used for checking its similarity with those of
	other SFGs. Given the positional matches and the empirical
	expectation 
	for \gr\ emission from M83 due to the galaxy's star-forming activity, 
	we conclude that the \gr\ source is the 
	likely counterpart to m83. The detection thus adds another member to
	the group of approximately a dozen SFGs, whose \gr\ emissions mostly
	have a cosmic-ray origin.
\end{abstract}


\section{Introduction}

Among more than 6000 \gr\ sources detected with the Large Area Telescope (LAT)
on board {\it the Fermi Gamma-ray Space Telescope (Fermi)} in all sky, the 
dominant class is active galactic nuclei (AGN; \citealt{4fgl-dr3}), whose 
high-energy emission is mostly radiated from their jets. Non-active galaxies 
thus do not show such emission. However there are approximately a dozen of 
the galaxies, either 
within the local group or
nearby, have been detected at \grs\ \citep{aje+20,xi+20}. While there are 
complications in the
production of the \gr\ emissions observed from these galaxies, for example 
AGN possibly hiding in some of them \citep{pen+19} and the \gr\ emission of
the local-group galaxy M31 being considered consisting of different 
components \citep{li+16,pvp16,ack+17,kmc19,zim+22,xin+23},
\gr\ luminosities of most of them 
well correlate with infrared (IR) or radio 1.4\,GHz luminosities 
\citep{abd+10,ack+12,aje+20,xi+20}.
This correlation is considered as an indicator for the cosmic ray (CR) origin
of the gamma-ray emissions. Supernova remnants (SNRs) produce CRs as their 
shock fronts serve as particle accelerators (e.g., \citealt{byk+18}), and the 
accelerated
particles emit at radio frequencies through the synchrotron process and
at high energies through the proton-proton collisions and/or 
leptonic processes (i.e., bremsstrahlung or inverse Compton scattering;
e.g., \citealt{der86}). On the other hand, SNRs are the results of massive 
stars ($M\gtrsim 8\,M_{\sun}$) evolving to the end of their lives on
relatively short, $\sim 10^7$\,yr timescales, and their densities are 
thus closely related to star-formation of a galaxy.
Given that the IR luminosities are an indicator of star-formation rates
of a galaxy, a correlation between them and corresponding \gr\ luminosities is 
naturally expected
for star-forming galaxies (see, e.g., \citealt{dt05,ltq10,ack+12}, and 
references therein).

Along this expected correlation, efforts have been made to detect nearby
star-forming galaxies at \grs. Thus far, approximately a dozen of them
have been reported with the detection (see, e.g., \citealt{aje+20,xi+20}, and 
references therein). Here we report on the likely detection of another one,
the nearby galaxy M83 (also known as NGC~5236 or the Southern Pinwheel).
\begin{figure*}
   \includegraphics[width=0.33\textwidth]{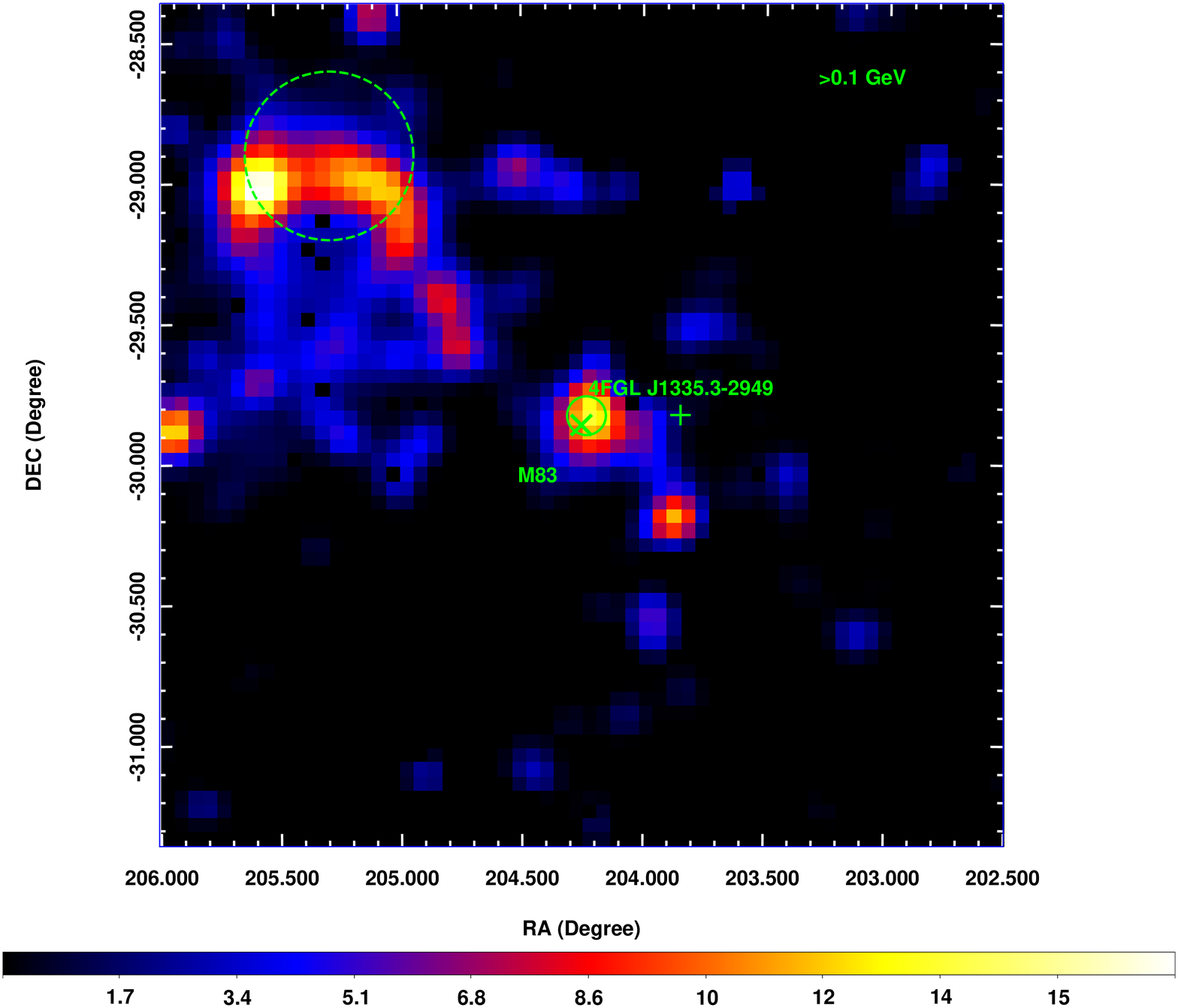}
   \includegraphics[width=0.33\textwidth]{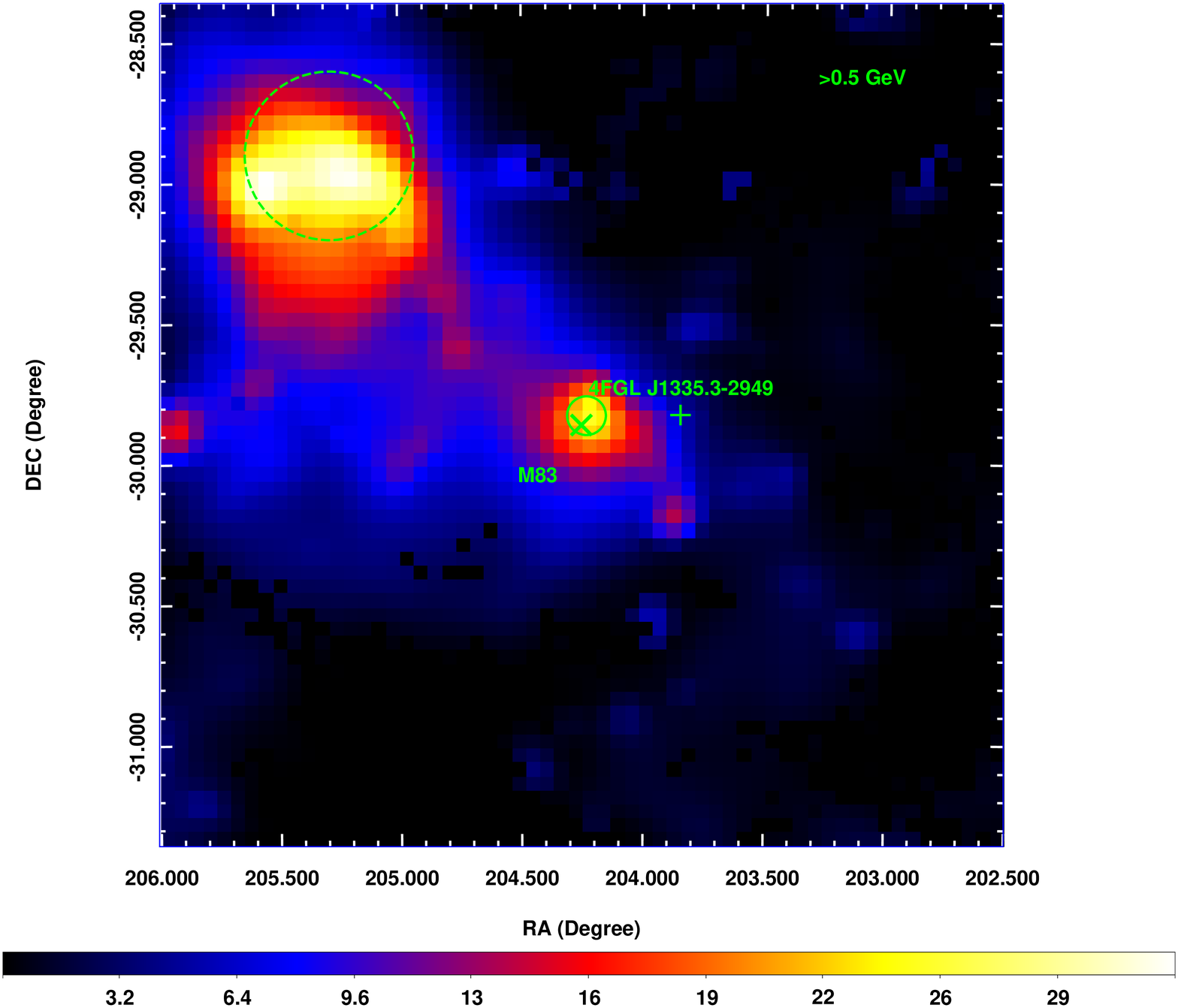}
   \includegraphics[width=0.33\textwidth]{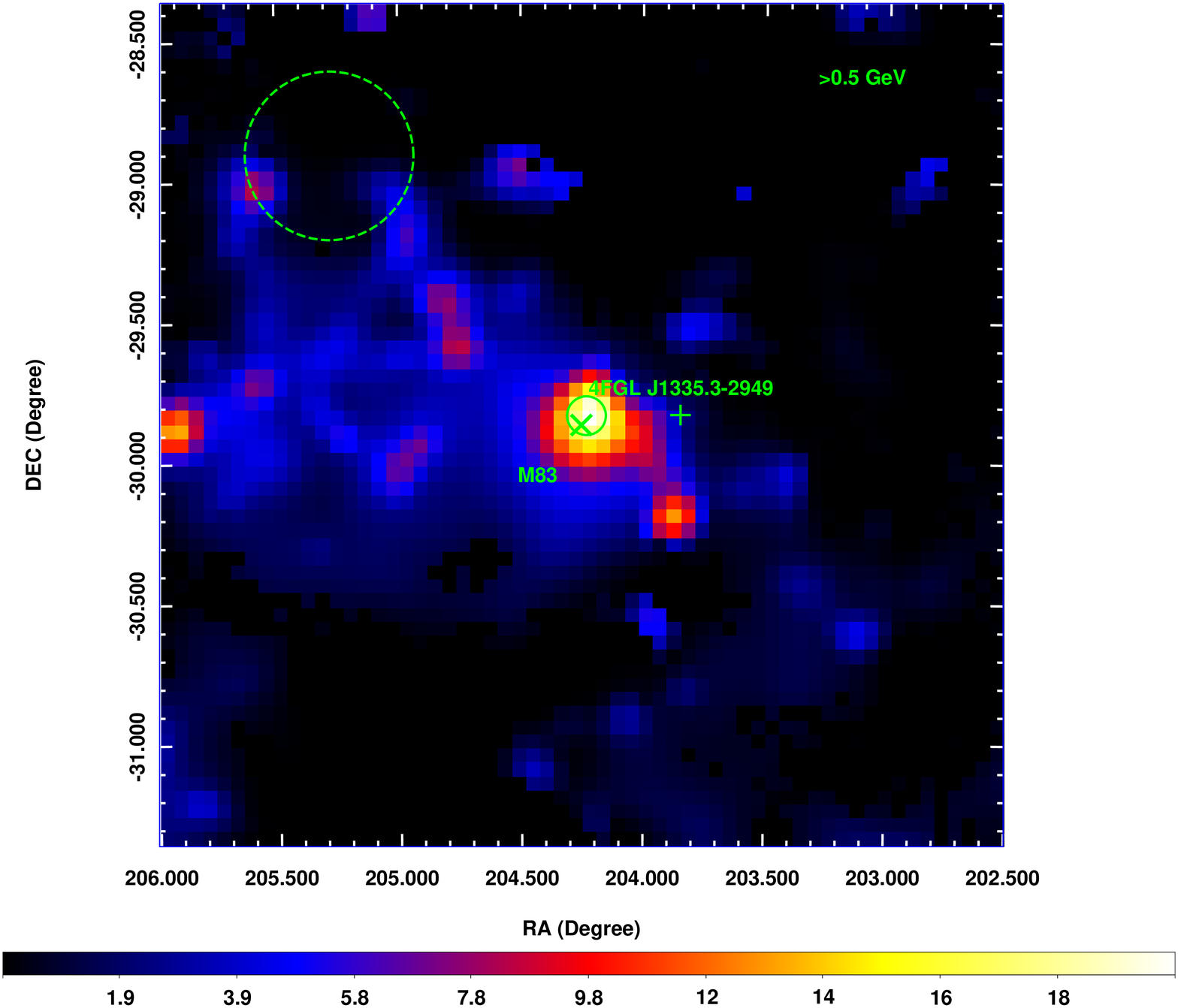}
   \caption{TS maps of a $3\arcdeg\times 3\arcdeg$ region centered at M83 
	in the energy ranges of 0.1--500\,GeV ({\it left} panel) 
	and 0.5--500\,GeV ({\it middle} panel). The only 4FGL-DR3 catalog 
	source in the region, a blazar candidate (marked by green pluses), 
	is removed in the maps. The green crosses mark the position of M83, 
	which is within
	the 2$\sigma$ error circle (marked by the green circles) determined
	for the residual excess emission at the location. There is also weak
	excess emission (whose 2$\sigma$ error circle is marked by green dashed
	circles) NE to the center. {\it Right} panel: the same as 
	the middle panel, but the NE source is removed as a point 
	source.
	For each TS map, the image scale is 0\fdg05\,pixel$^{-1}$, and the 
	color bar indicates the TS value range.  }
   \label{fig:tsmap}
\end{figure*}

M83 is often referred to as a grand-design spiral galaxy. It is nearly 
face-on to us ($i\sim 25\arcdeg$, \citealt{sof+99}), at a distance 
of 4.61\,Mpc \citep{sah+06}. There have been
extensively studies of the galaxy over the whole wavelength range. Its star
formation is relatively active, having a total star-formation rate of
5\,$M_{\sun}$\,yr$^{-1}$ \citep{ken98}. Thus it was listed in 
the star-forming galaxy sample selected by \citet{ack+12},
to be searched for CR-induced \gr\ emission.
A flux upper limit was reported in \citet{ack+12}, while only 3 years of
the \fermi-LAT data were used.

Now with 14 years of the data having been collected with \fermi-LAT, we 
conducted a search for M83's \gr\ emission.
We found a likely counterpart and report the results.
Below we describe the details of our analysis and provide the corresponding
results in Section~\ref{sec:ar}. The results are discussed in 
Section~\ref{sec:dis}.

\section{Analysis and Results}
\label{sec:ar}

\subsection{\fermi-LAT Data and Source Model}
\label{subsec:data}

We selected 0.1--500\,GeV LAT events from the updated \textit{Fermi} Pass 8 
database in a region of interest (RoI) that has a size
of 20$^{\circ}$ $\times$ 20$^{\circ}$ and the center at the central
position of M83. Since the galaxy appears to have a size $\sim 12\arcmin$ in
the sky, which is not resolvable in the LAT data due to its large point 
spread function (PSF), we treat M83 as a point source through this paper.
The time period of the LAT data was
from 2008-08-04 15:43:39 (UTC) to 2022-09-26 23:16:35 (UTC), slightly 
more than 14 yrs.
The \textit{CLEAN} event class was used.
We included the events with zenith angles less than 90\,deg
and excluded the events with quality flags of `bad'. 
Both these are recommended by the LAT team\footnote{\footnotesize http://fermi.gsfc.nasa.gov/ssc/data/analysis/scitools/}.

We constructed the source model by including all sources within 20 deg
from M83.  The positions and the spectral parameters of these sources
are provided in the \textit{Fermi} LAT 12-year source catalog 
(4FGL-DR3; \citealt{4fgl-dr3}).
We set the spectral parameters of the sources 
within 5 deg from M83 free, and froze
the other parameters at their catalog values. 
The spectral model gll\_iem\_v07.fit was used for the Galactic diffuse 
emission, and the spectral file iso\_P8R3\_CLEAN\_V3\_v1.txt for 
the extragalactic diffuse emission. The normalizations of these two diffuse 
components were set free in the following analyses.
\begin{table}
	\begin{center}
\caption{Likelihood analysis results with the PL, PLEC, and LP models}
	\label{tab:model}
	\begin{tabular}{lcccc}
	\hline
	Model & Best-fit parameters & $\log(L)$ & TS \\
	\hline
	PL  & $\Gamma = 2.0 \pm 0.2$  & 265522.5 & 14 \\
	PLEC & $\Gamma = 0.00 \pm 0.03$ & 265526.9 & 22\\
		& $E_c = 2.8 \pm 0.8$ &     &\\
	LP  & $\alpha = 0.00 \pm 0.06$ & 265526.1 & 20 \\
		& $\beta = 0.67 \pm 0.14$ & & \\
\hline
\end{tabular}
\end{center}
\end{table}

\subsection{Likelihood Analysis}
\label{sec:mla}

\begin{figure*}
	\centering
   \includegraphics[width=0.42\textwidth]{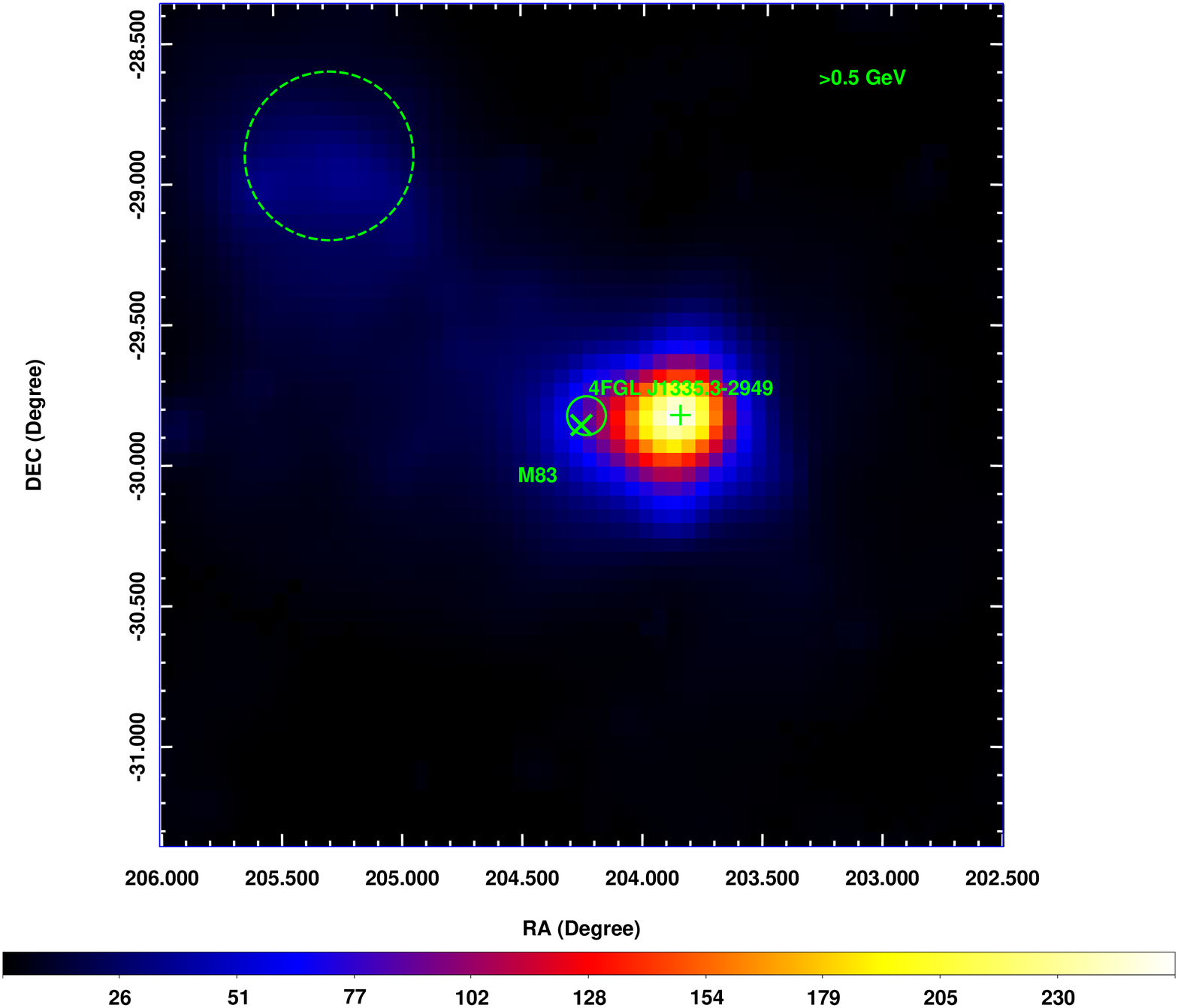}
   \includegraphics[width=0.42\textwidth]{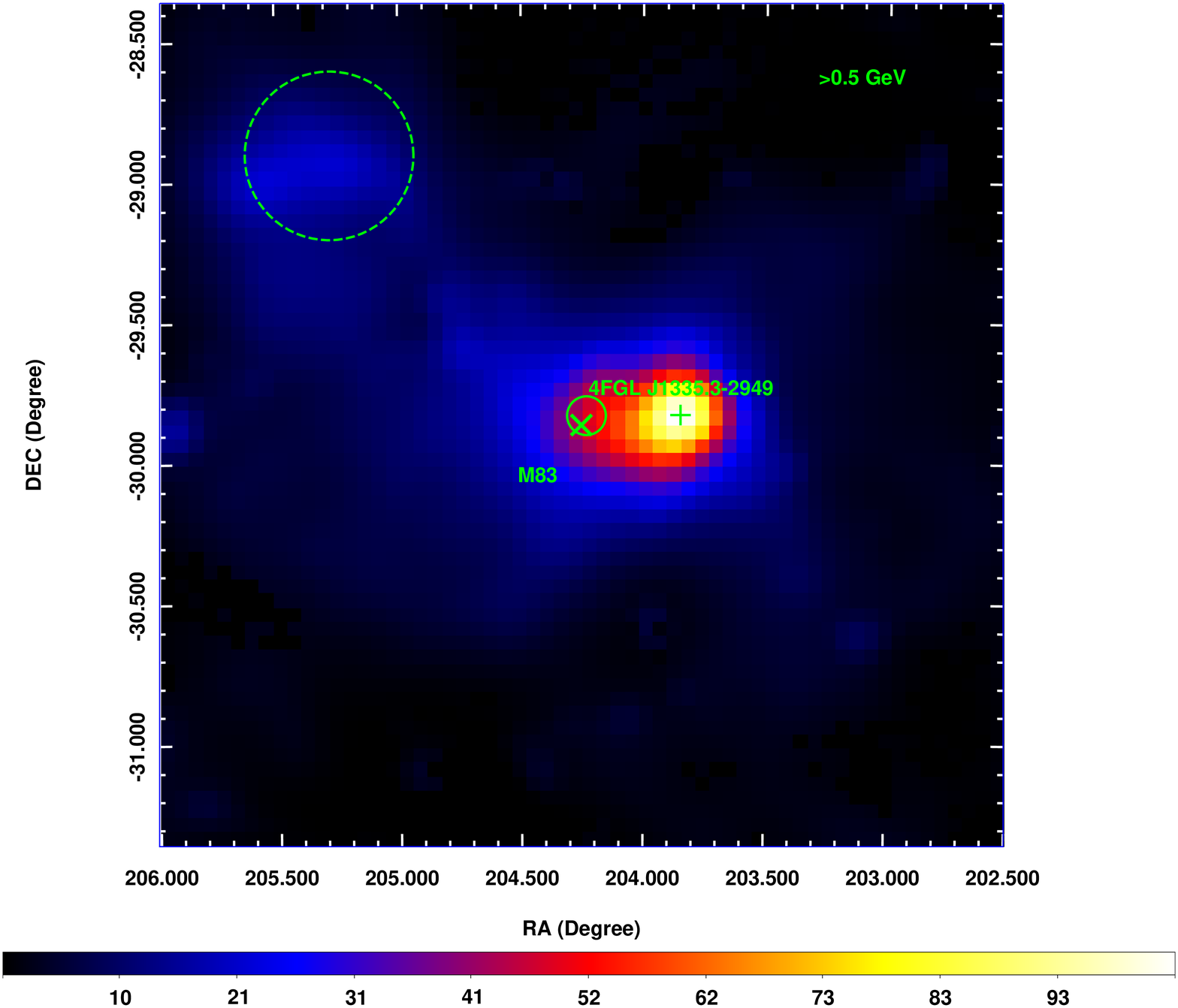}
	\caption{Same as the middle panel of Figure~\ref{fig:tsmap}, but
	the blazar candidate is kept in the maps. The {\it left} panel TS map is
	calculated from the whole LAT data, and the {\it right} panel one 
	from the time period of MJD~55042--59000
	(see Section~\ref{subsec:near} and Figure~\ref{fig:lc}).  }
   \label{fig:tsmap-quie}
\end{figure*}

We performed the standard binned likelihood analysis to the whole data in
0.1--500\,GeV and 
updated the parameter values for the sources within 5\,deg from M83. 
With the obtained results, we calculated a 0.1--500\,GeV residual 
Test Statistic (TS) map of a $\mathrm{3^{o}\times3^{o}}$ region centered at 
M83. The field is rather clean, with only one catalog source
(4FGL~J1335.3$-$2949; see Section~\ref{subsec:near}) in the region.
This source was included in the source model and removed in the TS map, 
which is shown in the left panel of Figure~\ref{fig:tsmap}. As can be seen, 
there is 
weak excess emission at the position of M83. The maximum TS value is 
approximately 14, corresponding to a $\sim$3.7$\sigma$ detection significance. 
Because in the analysis below (Section~\ref{sec:sa} and 
Table~\ref{tab:spectra}) we have found that the low energy range 
for significant detection is $\gtrsim$0.5\,GeV, 
we also calculated the 0.5--500 GeV residual TS map (shown in
the middle panel of Figure~\ref{fig:tsmap}).  The maximum TS value of  
the excess emission at M83 is improved to be $\sim$25, now at a 
$\sim$5$\sigma$ detection significance. We ran \textit{gtfindsrc} in 
the {\tt Fermitools} to the 0.5--500 GeV data to determine the position,
and obtained R.A.=204\fdg23, Decl.=$-$29\fdg83 (equinox J2000.0), with 
a 1$\sigma$ nominal uncertainty of 0\fdg04. M83 is 0\fdg04 away from 
this position and thus within the 1$\sigma$ error circle.

We then added this new source, the possible counterpart to M83, in the source 
model as a point source and repeated the likelihood analysis in 0.1--500\,GeV.
Given the low TS value of the source, we considered three models to fit its 
emission.  One is a simple power law (PL), $dN/dE=N_{0}E^{-\Gamma}$, 
where $\Gamma$ is the photon index, and the other two are a PL with 
an exponential cutoff (PLEC), $dN/dE=N_{0}E^{-\Gamma}\exp(-E/E_c)$, 
where $E_{c}$ is the cutoff energy, and a Log-Parabola (LP) function,
$dN/dE = N_{0}(E/E_{b})^{-[\alpha + \beta\log(E/E_{b})]}$, where $E_b$ is
a scale parameter and was fixed at 1\,GeV in our analysis.
The results obtained with the three models are given in Table~\ref{tab:model}.
The PLEC model provided the largest TS value, $\simeq$22. By comparing
the log-likelihood values, that is 
$\sqrt{-2 log(L_{i}/L_{j})}$, where $L_{i/j}$ are the maximum likelihood 
values from model $i$ and $j$, 
the PLEC and LP models are found to be more favored than the PL at 
$\sim$3.0$\sigma$ and $\sim$2.7$\sigma$ significances respectively.
Although \gr\ emissions of most star-forming galaxies can be well described 
with a PL model (or a LP model; e.g., \citealt{aje+20}), below we adopted 
the PLEC model because of the largest TS value resulting from it. The 
corresponding 0.1--500 GeV photon flux for the source was 
$F_{0.1-500}\simeq 1.2\pm0.4\times10^{-10}$\,photon\,cm$^{-2}$\,s$^{-1}$.
\begin{figure*}
   \centering
   \includegraphics[width=0.6\textwidth]{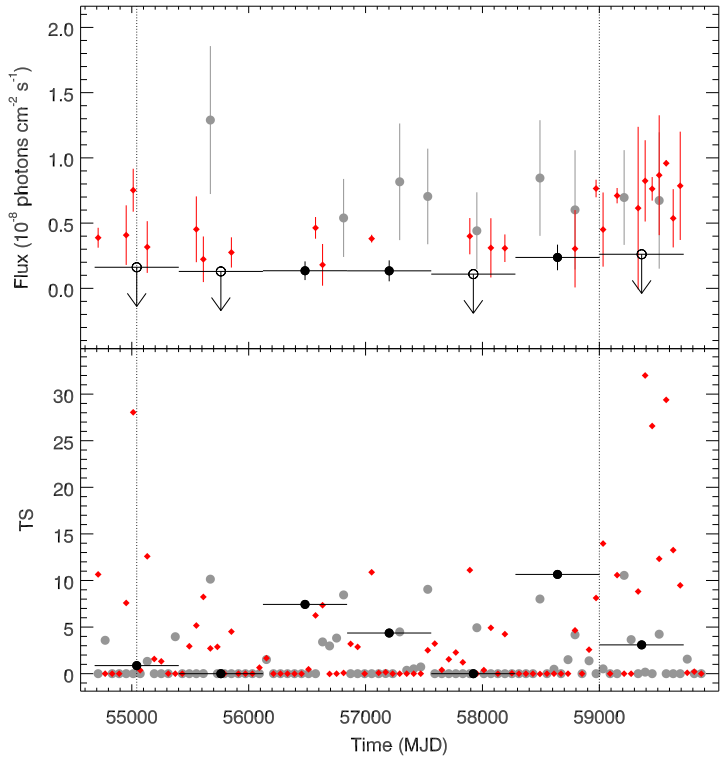}
	\caption{60-day binned light curves ({\it upper}) and TS curves
	({\it bottom}) of the blazar candidate (red) and the source at M83
	(gray) in 0.1--500\,GeV. 
	For the latter, 2-yr binned light curve and TS curve (black 
	data points) are also shown.
	Fluxes with TS$\geq$4 are kept in the light 
	curves. Two dotted lines mark the time period of MJD~55042--59000,
	during which no obvious flares are seen from the blazar candidate.
}
   \label{fig:lc}
\end{figure*}

\subsection{Analysis for Nearby Sources} 
\label{subsec:near}

Since the emission at the position of M83 was weak, we conducted extra-checks
to ensure the detection. First as shown in the TS maps (Figure~\ref{fig:tsmap}),
another excess emission is present, which is north-east (NE) to M83. Although
it does not appear like a point source, we ran \textit{gtfindsrc} to 
the 0.5--500\,GeV data, and obtained a position of R.A.=205\fdg3, 
Decl.=$-$28\fdg9 (equinox J2000.0), with a 1$\sigma$ nominal uncertainty 
of 0\fdg2. This position is 1.3\,deg away from M83, nearly outside of 
the 68\% containment angle of the LAT PSF in $>$0.5 GeV 
band\footnote{\footnotesize https://www.slac.stanford.edu/exp/glast/groups/canda/lat\_Performance.htm}. 
We added this source in the source model and performed the likelihood analysis.
The source could be totally removed in the TS map (shown in the right panel 
of Figure~\ref{fig:tsmap}), and the results for the emission at M83 were 
nearly the same as above.
These suggest that the NE excess emission did not affect our analysis results 
for the source at M83.
 
Second, the catalog source, 4FGL J1335.3$-$2949, is located very close 
to M83 (see Figure~\ref{fig:tsmap}).
It has an angular separation of 0\fdg35 from M83 and its
1$\sigma$ positional uncertainty is 0\fdg02, given in 
4FGL-DR3 \citep{4fgl-dr3}. Thus this source and M83's source are 
outside of the error circle of 
each other. Because it is relatively bright, 
it could contaminate the detection of the source at M83. 
We calculated the 0.5--500 GeV TS map with the sources (including the NE one) 
in the $\mathrm{3^{o}\times3^{o}}$ region kept. 4FGL~J1335.3$-$2949 is clearly 
seen as the brightest one (the TS value was $\sim$206)
in the field (left panel of Figure~\ref{fig:tsmap-quie}). 

This source is a blazar candidate but did not show significant \gr\ 
variations in 12 years of the \fermi-LAT data \citep{4fgl-dr3}. 
We extracted its 0.1--500 GeV light curve by setting 60-day time bins 
and performing the likelihood analysis to each time-bin data. 
In the extraction, only the normalization parameters of the sources 
within 5\,deg from M83 were set free. As a check, we also extracted a
light curve of the source at M83 with the same setup. 
The resulting light curves and TS curves for the two sources are shown
in Figure~\ref{fig:lc}, for which only flux data points with 
TS$\geq$4 were kept in the light curves. 
As can be seen, this blazar candidate had a high TS value in the beginning 
of the data at $\sim$MJD~55000 and likely showed a flaring event after
MJD~59000 (note that the latter part was not covered in the LAT 12-yr source
catalog). The source at M83 did not show any obvious variations.

To reduce any contamination possibly caused by the variations of 
4FGL~J1335.3$-$2949, we selected the LAT data during the time period 
of MJD~55042--59000 (marked as dotted lines in Figure~\ref{fig:lc}).
The 0.5--500 GeV TS map of this time period was calculated and 
is shown in the right panel of Figure~\ref{fig:tsmap-quie}. 
The sources in the TS-map region were kept.
The TS value for the blazar candidate is reduced to $\sim$100, and
the source at M83, appearing as an extension of the former to the east 
direction, is revealed. Thus the flaring activity of the blazar candidate
could affect our analysis and weaken the visibility of the source at M83,
but when the flares were filtered out, the detection of this source was
proved to be true.
\begin{figure}
   \centering
   \includegraphics[width=0.46\textwidth]{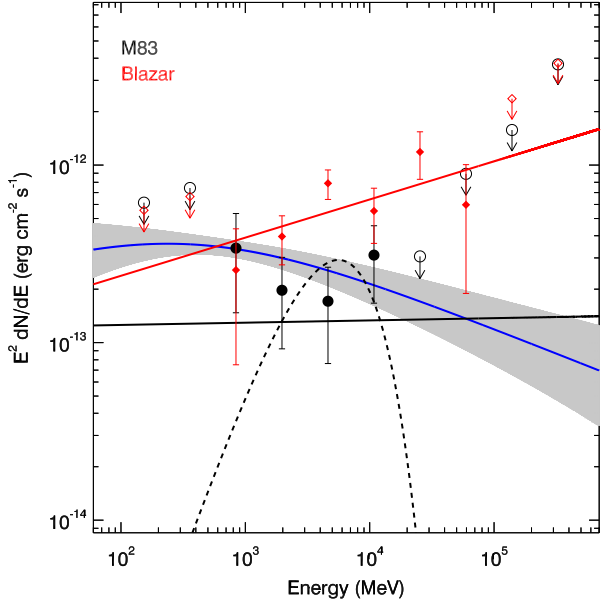}
	\caption{\gr\ spectra obtained for M83's source (black circles) and 
	the blazar candidate (red diamonds), for which the model fits from 
	the likelihood analysis are also shown for comparison, where
	the black dashed and solid curves are the PLEC and PL model fits 
	(cf., Section~\ref{sec:mla}), respectively, for the former source 
	and the red line is the PL model fit for the latter one.
	The model fit to the 
	normalized spectra of 9 star-forming galaxies determined by
	\citet{aje+20} is shown as the blue curve, while its 1$\sigma$
	uncertainty range is marked by the gray region (here  
	this model-fit curve is simply aligned with the first 
	flux measurement of M83's source).
}
   \label{fig:spectra}
\end{figure}

\subsection{Spectral Analysis}
\label{sec:sa}

We extracted the \gr\ spectrum of the source at M83 by performing maximum 
likelihood analysis to the LAT data in 10 evenly 
divided energy bands in logarithm from 0.1--500 GeV. 
In the extraction, the spectral normalizations of the sources within 5\,deg 
from M83 were set as free parameters, while all the other parameters 
of the sources were fixed at the values obtained from the above maximum 
likelihood analysis. The emission was set to be a PL with $\Gamma$ fixed to 2.
For the result, we kept only spectral data points when 
TS$\geq 4$ ($\geq$2$\sigma$ significance) and derived 95\% flux upper 
limits otherwise, where for the latter a Bayesian approch implemented
in the Python tool {\tt IntegralUpperLimit} (provided in the \fermi\ Science
tools) was used. 
The obtained spectrum is plotted as black points in Figure~\ref{fig:spectra}, 
and the flux and TS values of the spectral data points are provided in 
Table~\ref{tab:spectra}. In the energy band of 0.2--0.5 GeV, the TS value 
is 5, but the flux has a large uncertainty, 
0.42$\pm$0.62$\times$10$^{-12}$\,erg\,cm$^{-2}$\,s$^{-1}$, likely caused by
the contamination of
the nearby blazar candidate (because of large containment angles of the LAT
PSF at the low energies; see also Figure~\ref{fig:spectra}). Thus for
this data point, we report
an upper limit instead. In addition, the spectrum of the nearby blazar candidate
was obtained with the same setup, and is also shown in Figure~\ref{fig:spectra}
for comparison.

\subsection{Variability Analysis}

We checked the source at M83 for any long-term variability 
in 0.1--500\,GeV by calculating the variability 
index TS$_{var}$ \citep{nol+12}. We set 87 time bins with each
bin consisting of 60-day data and derived the fluxes or flux upper limits
for the source (i.e., shown in Figure~\ref{fig:lc}).
If the emission is constant, 
TS$_{var}$ would be distributed as a $\chi^{2}$ distribution with 86 degrees 
of freedom. 
A variable source would be identified if TS$_{var}$ is larger than 119.4 
(at a 99\% confidence level).
The computed TS$_{var}$ for the source is 74.8, lower than the 
threshold value. Since this source is faint, we also
constructed its 2-yr binned light curve (see Figure~\ref{fig:lc}) and checked 
for variability. For 7 time bins (i.e., 6 degrees of freedom), 
TS$_{var}>16.8$ is required for a variable source. We obtained 
TS$_{var}\simeq 11.3$. Thus there were no significant long-term variations
found for this source.
\begin{table}
\begin{center}
\caption{Flux Measurements}
\label{tab:spectra}
\begin{tabular}{lccccc}
\hline
Band & $G_{M83}/10^{-12}$ & TS \\
	(GeV) & (erg cm$^{-2}$ s$^{-1}$) & \\\hline
0.15 (0.1--0.2) & 0.61 & 0 \\
0.36 (0.2--0.5) & 0.74 & 5 \\
0.84 (0.5--1.3) & 0.34$\pm$0.19 & 7 \\
1.97 (1.3--3.0) & 0.20$\pm$0.10 & 6 \\
4.62 (3.0--7.1) & 0.17$\pm$0.09 & 5 \\
10.83 (7.1--16.6) & 0.31$\pm$0.14 & 13 \\
25.37 (16.6--38.8) & 0.31 & 0 \\
59.46 (38.8--91.0) & 0.89 & 0 \\
139.36 (91.0--213.3) & 1.58 & 0 \\
326.60 (213.3--500.0) & 3.69 & 0 \\
\hline
\end{tabular}
\\
\footnotesize{Note: Fluxes without uncertainties are the 95$\%$ upper limits.}
\end{center}
\end{table}

\section{Discussion}
\label{sec:dis}

By analyzing the \fermi-LAT data for the M83 region, we have found a faint
\gr\ source at the position of M83. The source's emission is
preferably described with a curved-function model (a PLEC or a LP). Although
the detection significance for the source is low, only $\sim 4.7\sigma$, and
a nearby blazar candidate complicates the detection (particularly at the
$\lesssim$0.5\,GeV low energies), our 
detailed analysis has ensured the existence of the source. Given the high
positional coincidence between the \gr\ source and M83 and the expectation
for \gr\ emission from this star-forming galaxy, we conclude that this 
\gr\ source is the likely counterpart to M83.

At a distance of 4.61\,Mpc \citep{sah+06}, the \gr\ luminosity 
(in 0.1--500\,GeV) of the source is 
$\sim 1.4\pm0.5 \times 10^{39}$\,erg\,s$^{-1}$, higher than that of
the Milky way \citep{ack+12} and lower than that of NGC~253
(a spiral galaxy with a central starburst region; \citealt{abd+10a,aje+20}).
The 2--1000\,$\mu$m IR luminosity of M83 is approximately
$8.7\times 10^{43}$\,erg\,s$^{-1}$ (\citealt{ack+12}; note that a source
distance of 3.7\,Mpc was used in \citealt{ack+12}). 
If we use the parameters obtained in \citet{aje+20} for the \gr--IR
luminosity correlation, the predicted
\gr\ luminosity for M83 would be $\sim 7.9\times 10^{39}$\,erg\,s$^{-1}$.
Further considering a dispersion of 0.3 (\gr--luminosity residuals with 
respect to
the correlation line in log space; \citealt{ack+12,aje+20}), the luminosity
range would be $\sim 4.0$--$16\times 10^{39}$\,erg\,s$^{-1}$. Thus the observed
luminosity is slightly below the correlation line but can be considered to
be consistent with it, given significant uncertainties such as on 
the distance and IR properties (\citealt{aje+20} and references therein).

Because of the weak detection of M83, its \gr\ spectrum only contains 4
data points (Figure~\ref{fig:spectra}). The model fits we obtained in
Section~\ref{sec:mla}, either the PLEC or the LP, appear to be highly curved
(for example, $\beta\approx 0.67$ in the LP model), which 
are different from those of the other star-forming galaxies. The latter
are mostly
described with a PL without significant curvature required
(e.g., \citealt{aje+20}). The difference could be due to the weak 
detection causing the emission property not well determined,
and remains to be resolved when more LAT data for M83 are collected.
\citet{aje+20} have normalized the \gr\ spectra of 9 star-forming
galaxies by simply scaling the spectra to a common value at 1\,GeV, and
obtained a best-fit model that is in the form of
a smoothly broken power law. In Figure~\ref{fig:spectra}, we show this best fit
by aligning it with the first flux measurement of M83's spectrum,
since the energy range of the data point is 0.5--1.3\,GeV,
approximately at 1\,GeV.
As can be seen, the spectrum 
of M83 is approximately
consistent with the best fit. Given this and the \gr\ luminosity indicated
above being approximately in the right range, the \gr\ emission of M83 likely 
has the same
origin as the other nearby galaxies, arising from CRs and related to the 
star-formation activity \citep{ack+12}. Thus M83 is likely another member of 
this gourp of the \gr--emitting,
star-forming galaxies. Hopefully with more LAT data collected in the future,
both the detection significance and the quality of the \gr\ spectrum will be 
improved, helping provide more information for the high-energy properties 
of M83.

\begin{acknowledgements}
This research is supported by the Original Innovation Program of the Chinese 
	Academy of Sciences (E085021002), 
	the Basic Research Program of Yunnan Province
No. 202201AS070005, and the National Natural Science Foundation of China
No. 12273033.
\end{acknowledgements}

\bibliographystyle{aasjournal}
\bibliography{gal}

\end{document}